\documentclass{PoS}

\leftline{\parbox{20cm}{\large DESY 13 - 190, Edinburgh 2013/25}} 

\title{Perturbatively improving renormalization constants}

\ShortTitle{Perturbatively improving renormalization constants}

\author{M.~Constantinou$^a$, M.~Costa$^a$,  M.~G\"ockeler$^{b}$, R.~Horsley$^{c}$,  
H.~Panagopoulos$^a$,\newline
\speaker{H.~Perlt}$^d$,
   P.~E.~L.~Rakow$^e$, G.~Schierholz$^f$,
   A.~Schiller$^d$\\
\llap{$^a$}Department of Physics, University of Cyprus, Nicosia, CY-1678, Cyprus\\
\llap{$^b$}Institut f\"ur Theoretische Physik, Universit\"at Regensburg,
93040 Regensburg, Germany\\
\llap{$^c$}School of Physics, University of Edinburgh,
Edinburgh EH9 3JZ, UK\\
\llap{$^d$}Institut f\"ur Theoretische Physik, Universit\"at Leipzig,
04109 Leipzig, Germany\\
\llap{$^e$}Theoretical Physics Division, Department of Mathematical Sciences,
University of Liverpool, Liverpool L69 3BX, UK\\
\llap{$^f$}Deutsches Elektronen-Synchrotron DESY,
22603 Hamburg, Germany\\
\\
{\rm E-mail}: perlt@itp.uni-leipzig.de}

\abstract{
Renormalization factors relate the observables obtained on the lattice to their 
measured counterparts in the continuum in a suitable renormalization scheme. 
They have to be computed very precisely which requires
 a careful treatment of lattice artifacts. 
In this work we present a method to suppress these artifacts 
by subtracting one-loop
contributions proportional to the square of
the lattice spacing  calculated in lattice perturbation 
theory.}

\FullConference{31st International Symposium on Lattice Field Theory - LATTICE 2013\\
		July 29 - August 3, 2013\\
		Mainz, Germany}

\begin{document}

\section{Introduction}

Renormalization factors in lattice Quantum Chromodynamics (QCD) relate 
observables computed on finite lattices to their
continuum counterparts in specific renormalization schemes. Therefore,
their determination should be as precise as possible in order to allow for
a reliable comparison with experimental results. 
A widely used method to calculate these factors is the so-called
Rome-Southampton method~\cite{Martinelli:1994ty} (utilizing the RI-MOM scheme).
Like (almost) all quantities evaluated in lattice QCD also renormalization
factors computed in this non-perturbative scheme suffer from discretization effects. 
In this paper we describe a method to suppress
these lattice artifacts using a subtraction procedure based
on perturbation theory. It has been published in Ref.~\cite{Constantinou:2013ada} and
the reader is referred for all details to this reference.

In a recent paper of the QCDSF/UKQCD collaboration~\cite{Gockeler:2010yr}  a 
comprehensive discussion and comparison of perturbative and 
nonperturbative renormalization have been given. 
It was shown that a subtraction of the complete lattice artifacts in one-loop lattice 
perturbation theory improves the results for the Z factors
significantly. While being very effective this procedure is rather involved and
not suited as a general method for more complex operators, especially
for operators with more than one covariant derivative,
and complicated lattice actions.
An alternative approach can be based on the subtraction of
one-loop terms of order $a^2$, with $a$ 
being the lattice spacing. The computation of those
terms has been developed by the authors of Ref.~\cite{Constantinou:2009tr}
and applied to various operators for different actions.

We study the flavor-nonsinglet quark-antiquark operators given in Table~\ref{OpTab}. 
 \begin{table*}[!htb] 
   \begin{center}
 \begin{tabular}{|c|c|c|l|}\hline
   Operator (multiplet)  & Notation & Representation & Operator basis \\ \hline
   $\bar{u}\,d$  & $\mathcal{O}^{S}$        & $\tau_1^{(1)}$ &   $\mathcal{O}^{S}$    \\
   $\bar{u}\,\gamma_\mu\,d$  & $\mathcal{O}_\mu^{V}$        & $\tau_1^{(4)}$ &   $\mathcal{O}_1^{V}, \mathcal{O}_2^{V}, \mathcal{O}_3^{V}, \mathcal{O}_4^{V}$    \\
   $\bar{u}\,\gamma_\mu\gamma_5\,d$  & $\mathcal{O}_\mu^{A}$        & $\tau_4^{(4)}$ &   $\mathcal{O}_1^{A}, \mathcal{O}_2^{A}, \mathcal{O}_3^{A}, \mathcal{O}_4^{A}$    \\
   $\bar{u}\,\sigma_{\mu\nu}\,d$  & $\mathcal{O}_{\mu\nu}^{T}$        & $\tau_1^{(6)}$ &   $\mathcal{O}_{12}^{T}, \mathcal{O}_{13}^{T}, \mathcal{O}_{14}^{T}, \mathcal{O}_{23}^{T}, \mathcal{O}_{24}^{T}, \mathcal{O}_{34}^{T}$    \\ 
   $\bar{u}\,\gamma_\mu \stackrel{\leftrightarrow}{D_\nu}\,d$  & $\mathcal{O}_{\mu\nu}\to\mathcal{O}^{v_{2,a}}$        & $\tau_3^{(6)}$ &   $\mathcal{O}_{\{12\}}, \mathcal{O}_{\{13\}}, \mathcal{O}_{\{14\}}, \mathcal{O}_{\{23\}}, \mathcal{O}_{\{24\}}, \mathcal{O}_{\{34\}}$    \\
   $\bar{u}\,\gamma_\mu \stackrel{\leftrightarrow}{D_\nu}\,d$  & $\mathcal{O}_{\mu\nu}\to\mathcal{O}^{v_{2,b}}$        & $\tau_1^{(3)}$ &   $1/2(\mathcal{O}_{11}+ \mathcal{O}_{22}- \mathcal{O}_{33}- \mathcal{O}_{44})$,    \\ 
    & &  &   $1/\sqrt{2}(\mathcal{O}_{33}- \mathcal{O}_{44}), 1/\sqrt{2}(\mathcal{O}_{11}-\mathcal{O}_{22})$    \\ \hline
\end{tabular} 
   \end{center}
 \caption{\label{OpTab}Operators and their representations with respect to the hypercubic group 
 as investigated in the 
present paper. The symbol
$\{...\}$ means total symmetrization. A detailed group theoretical
discussion is given in~\cite{Gockeler:1996mu}.}
\end{table*}
The corresponding renormalization factors have been measured 
(and chirally extrapolated) at the lattice coupling $\beta=6/g^2=5.20, 5.25, 5.29$ and  $5.40$
using $N_f=2$ clover improved Wilson fermions with plaquette gauge 
action~\cite{Gockeler:2010yr}.
The Sommer scale $r_0$ is taken to be $r_0=0.501 \, {\rm fm}$ and the relation between the lattice spacing $a$
and  $\beta$ is given by $r_0/a =6.050\,(\beta=5.20), 6.603\,(\beta=5.25),
7.004\,(\beta=5.29)$ and $8.285\,(\beta=5.40)$~\cite{Bali:2012qs}.
This results in the corresponding values $a(\beta)=(0.42, 0.39, 0.36, 0.31) {\rm GeV}^{-1}$.
All results are computed in Landau gauge. The clover parameter $c_{SW}$
used in the perturbative calculation discussed below is set to its lowest order value $c_{SW}=1$.

\section{\label{sec:RGI}Renormalization group invariant operators}

We define a so-called RGI (renormalization group invariant) operator, which is independent
of scale $M$ and scheme $\mathcal{S}$, by~\cite{Gockeler:2010yr}
\begin{equation}
\mathcal{O}^{\rm RGI} = \Delta Z^{\mathcal{S}}(M)\, \mathcal{O}^{\mathcal{S}}(M) = Z^{\rm RGI}(a) \,\mathcal{O}_{\rm bare}\,
\label{RGI1}
\end{equation}
with
\begin{equation}
 \Delta Z^{\mathcal{S}}(M) = \left(2\beta_0 \frac{g^\mathcal{S}(M)^2}{16\,\pi^2}\right)^{-(\gamma_0/2\beta_0)}
\, {\rm exp}\left\{ \int_0^{g^{\mathcal{S}}(M)} dg'   \left( \frac{\gamma^{\mathcal{S}}(g')}{\beta^{\mathcal{S}}(g')}+
 \frac{\gamma_0}{\beta_0 g'} \right)  \right\}
 \label{RGI2}
\end{equation}
and the RGI renormalization constant (depending on the lattice spacing $a$ via the lattice coupling)
\begin{equation}
Z^{\rm RGI}(a) = \Delta Z^{\mathcal{S}}(M) \,\, Z^{\mathcal{S}}_{\rm bare}(M,a)\,.
\label{RGI3}
\end{equation}
Here $g^\mathcal{S}$, $\gamma^{\mathcal{S}}$ and $\beta^{\mathcal{S}}$ 
are the coupling constant, the anomalous dimension and the
$\beta$-function in scheme $\mathcal{S}$, respectively. Relations 
(\ref{RGI1}), (\ref{RGI2}) and (\ref{RGI3}) allow us to compute
the operator $\mathcal{O}$ in any scheme and at any scale we like,
once $Z^{\rm RGI}$ is known.
Ideally, $Z^{\rm RGI}$ depends only on the bare lattice coupling,
but not on the momentum $p$ which determines the scale via $M^2 = p^2$.
Computed on a lattice, however, it suffers from lattice artifacts,
e.g., it contains contributions proportional to $a^2p^2$,
$(a^2p^2)^2$ etc. For a 
precise determination it is essential to have these discretization 
errors under control.

As the ${\rm RI}^\prime$-MOM scheme is in general not $O(4)$-covariant even in the
continuum limit, it is not very suitable for computing the anomalous
dimensions needed in (\ref{RGI2}).
Therefore we use an intermediate scheme $\mathcal{S}$ with known 
anomalous dimensions and calculate $Z^{\rm RGI}$ as follows:
\begin{equation}
Z^{\rm RGI}(a) = \Delta Z^{\mathcal{S}}(M)\, Z^{\mathcal{S}}_{\rm RI^\prime-MOM}(M)
\, Z^{\rm RI'-MOM}_{\rm bare}(M,a)\,.
\label{RGI4}
\end{equation}
It turns out that a type of momentum subtraction scheme is a good
choice for $\mathcal S$ (for details see Ref.~\cite{Gockeler:2010yr}). 

\section{\label{sec:PTSUB2}Subtraction of order $a^2$ one-loop lattice artifacts}

The diagrammatic approach to compute the one-loop $a^2$ terms for the 
$Z$ factors of local and one-link operators has been developed
in Ref.~\cite{Constantinou:2009tr}. 
The general case of Wilson type improved fermions is discussed 
in~\cite{Alexandrou:2012mt}. We compute a common $Z$ factor for each
multiplet given in Table~\ref{OpTab}. 
As examples we give here (up to terms of higher order in the bare coupling $g^2$ and
$a^2$)
\begin{eqnarray}
Z_S &=& 1 + \frac{g^2\,C_F}{16 \pi^2}\,
\Bigg\{-23.3099 + 3\,\log(a^2S_2) \nonumber\\
&&+\, a^2\left[S_2\left(1.64089 - \frac{239}{240} \log (a^2{S_2})\right)+
     \frac{S_4}{S_2}\left(1.95104 - \frac{101}{120}\log (a^2{S_2})\right)\right]\Bigg\}\,,\\
Z_{v_{2,a}} &=&   1 + \frac{g^2\,C_F}{16 \pi^2}\,
\Bigg\{ 6.93831 -
      \frac{8}{3}\log (a^2 S_2) -
      \frac{2}{9}\frac{S_4}{(S_2)^2} \nonumber \\
&&+\,{a^2}\,\Bigg[
          S_2\,\left( -1.50680 +
            \frac{167}{180} \log (a^2S_2)\right)
\nonumber \\
 &&\hspace{0.3cm}
 +\frac{S_4}{S_2}\,\left( 2.63125 -
              \frac{197}{180}\log (a^2S_2) \right)-\frac{71}{540}\frac{{S_4}^2}{(S_2)^3} -
         \frac{82}{135}\frac{S_6}{(S_2)^2} \Bigg]\Bigg\}\,,
\end{eqnarray}
where we have introduced the notations $ S_n = \sum_{\lambda=1}^4 \, p_\lambda^n$
($p_\lambda$ being the momentum components) and $C_F=4/3$. 
The $Z$ factors are written generically as
\begin{equation}
 Z = 1 + \frac{g^2\,C_F}{16 \pi^2}\,Z_{\rm 1-loop}(p,a) + {a^2} g^2 Z^{(a^2)}_{\rm 1-loop}(p,a)\,.
\label{Za2corr}
\end{equation}
We emphasize that the numerical coefficients in the above expressions 
are either exact rationals or can be computed to a very high precision.

\begin{figure*}[!htb]
  \begin{center}
     \begin{tabular}{lcr}
        \includegraphics[scale=0.5,clip=true]
         {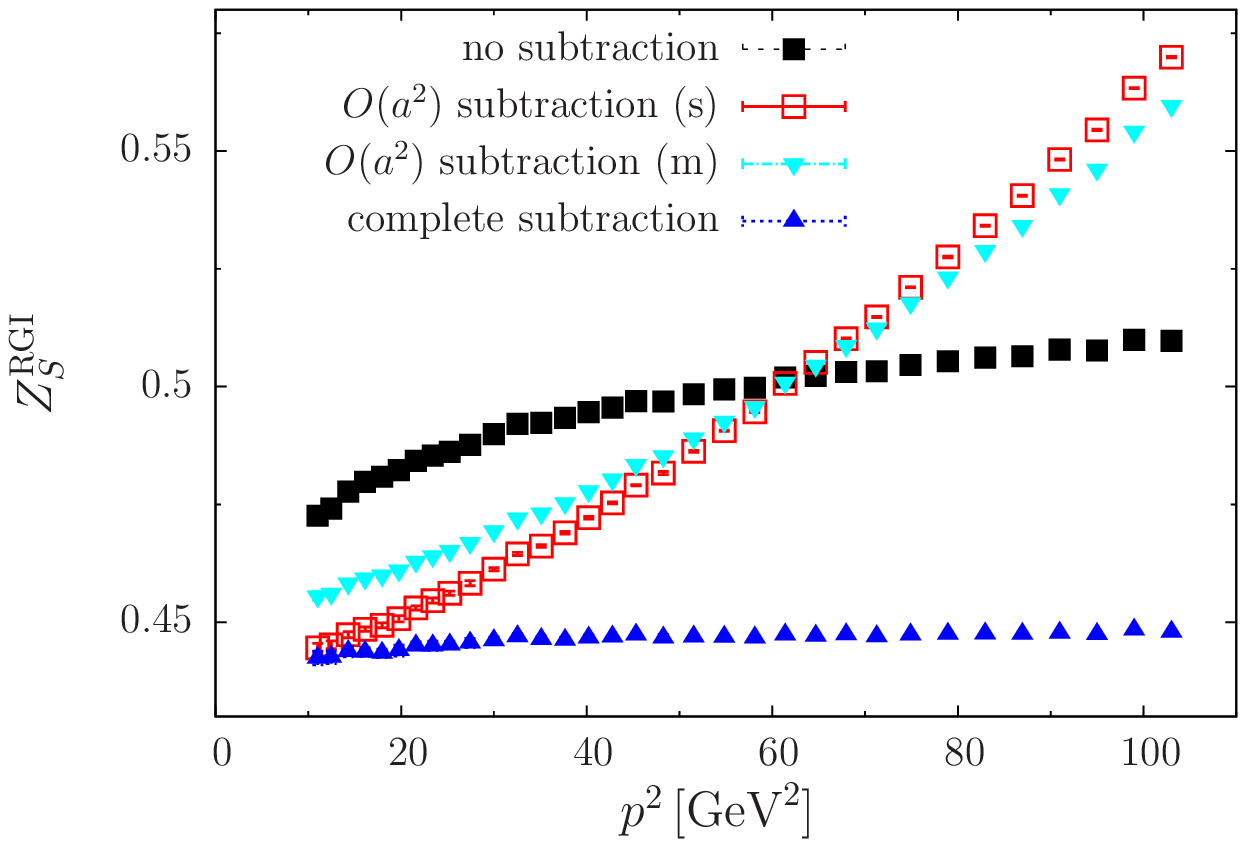}
&~\hspace{1cm}~&
       \includegraphics[scale=0.5,clip=true]
         {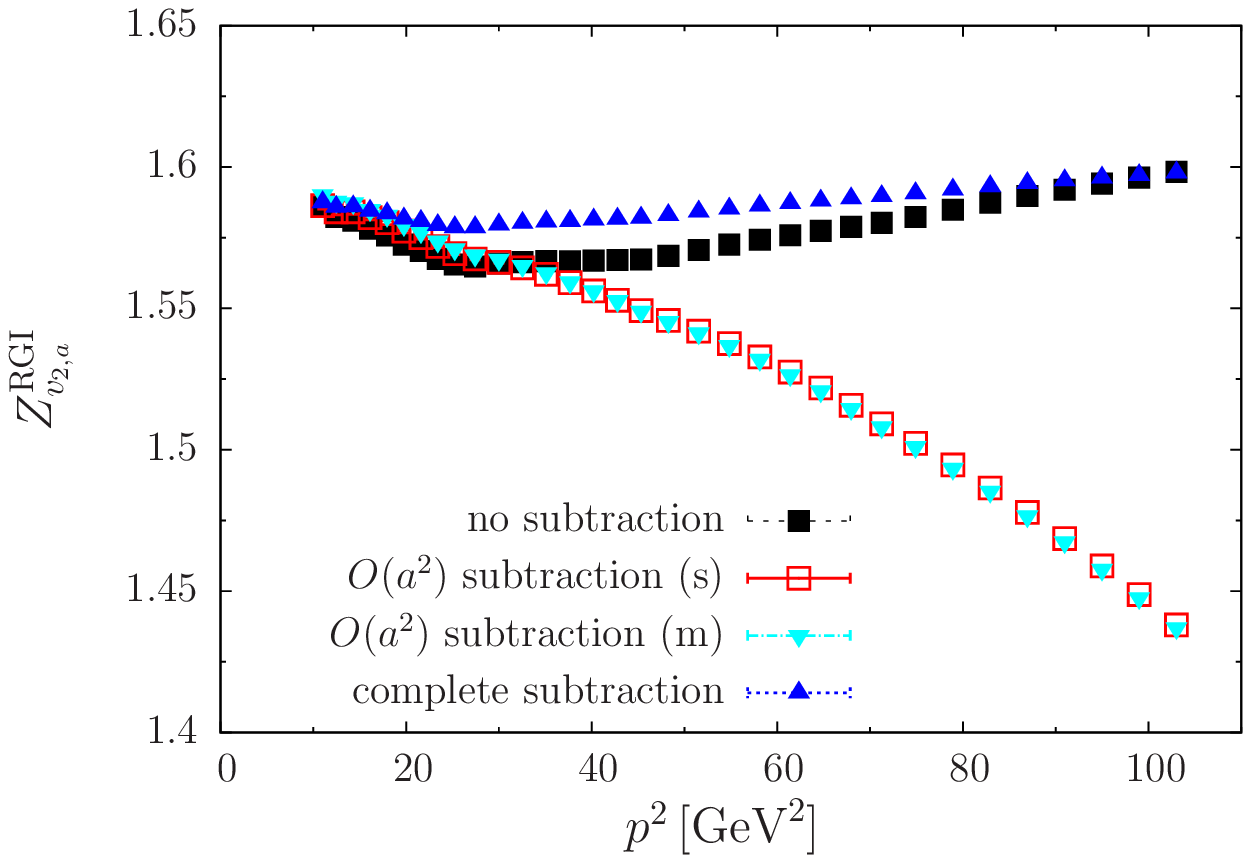}
     \end{tabular}
  \end{center}\vspace{-0.05cm}
  \caption[]{\label{fig:ZSZT}Unsubtracted and subtracted renormalization constants for the scalar operator $\mathcal{O}^S$ (left)
and the one-link operator $\mathcal{O}^{v_{2,a}}$ (right) at $\beta=5.40$, for $p^2 \gtrsim 10\, {\rm GeV}^2$
and $r_0\,\Lambda_{\rm \overline{MS}}=0.700$. The $a^2$ subtractions are of type 
({\bf s}) and ({\bf m}) with $g_\star=g^{}_{\rm B}$.}
\end{figure*}

Now we turn to the subtraction procedure applied to the Monte Carlo data 
$Z^{\rm RI'-MOM}_{\rm bare}(p,a)_{\rm MC}$. The subtraction of order $a^2$ 
terms is not unique - we have different possibilities. The only restriction 
is that in one-loop perturbation theory they should agree. We investigate the
following definitions (choices ({\bf s}) and ({\bf m})) of subtracted renormalization constants,
\begin{eqnarray}
Z^{\rm RI'-MOM}_{\rm bare}(p,a)_{\rm MC,sub,s} &=& Z^{\rm RI'-MOM}_{\rm bare}(p,a)_{\rm MC} - 
 {a^2} \, g_\star^2 \, Z^{(a^2)}_{\rm 1-loop}(p,a)\,, \label{eq:pts11} \\
Z^{\rm RI'-MOM}_{\rm bare}(p,a)_{\rm MC,sub,m} &=& Z^{\rm RI'-MOM}_{\rm bare}(p,a)_{\rm MC}\, \times
  \left(1 -{a^2} \, g_\star^2 \, Z^{(a^2)}_{\rm 1-loop}(p,a)\right)\,, \label{eq:pts12} 
\end{eqnarray}
where $g_\star^2$ can be chosen to be either  $g^2$ 
or the boosted coupling 
$g_{\rm B}^2=g^2/P(g) = g^2 + O(g^4)$, $P(g)$ being the measured plaquette at $\beta=6/g^2$.
The effect of the different subtractions is shown
in Fig.~\ref{fig:ZSZT}.
The complete one-loop subtraction leads to almost flat $Z^{\rm RGI}$, whereas 
the Z factors obtained from the one-loop $a^2$ subtractions show a significant
dependence on $p^2$ which has to be taken into account in the calculation.

We expect that $Z^{\rm RI'-MOM}_{\rm bare}(p,a)_{\rm MC,sub}$  
contains terms proportional to 
$a^{2n}$ ($n \geq 2$) even at order $g^2$, as well as the lattice
artifacts from higher orders in perturbation theory, constrained
only by hypercubic symmetry. Therefore, we parametrize the subtracted 
data for each $\beta$ in terms of the hypercubic invariants $S_n$ as follows
\begin{eqnarray}
 Z^\mathcal{S}_{\rm RI'-MOM}(p)\, &&\hspace{-3mm}
			Z^{\rm RI'-MOM}_{\rm bare}(p,a)_{\rm MC,sub} = 
  \frac{Z^{\rm RGI}(a)}{ \Delta Z^{\mathcal{S}}(p)\,\left[1+b_1\,(g^\mathcal{S})^8\right]} +
\label{struc1}
\\ 
&&\hspace{-1cm}
a^2 \left( c_1\,S_2 + c_2 \,\frac{S_4}{S_2} +  c_3 \,\frac{S_6}{(S_2)^2}\right)
 + a^4 \, \left( c_4\, (S_2)^2 + c_5\, S_4 \right)
 +a^6 \left(c_6\, (S_2)^3 + c_7\, S_4 \,S_2 + c_8 \,S_6\right) \,.\nonumber
 \nonumber
\end{eqnarray}
The parameter $b_1$ is introduced to compensate for the truncation errors 
in the expressions from continuum perturbation theory.
There are also further non-polynomial invariants at order
$a^4, a^6$, but their behavior is expected to be well
described by the invariants which have been included already.

Together
with the target parameter $Z^{\rm RGI}(a)$ we have ten parameters  for this general case.
In view of the limited number of data points for each single $\beta$ value
$(5.20$, $5.25$, $5.29$, $5.40)$ we apply the ansatz (\ref{struc1}) at several $\beta$ values simultaneously, 
assuming $\beta$-independent fit parameters $b_1$ and $c_i$.
The renormalization factors are influenced by the choice for $r_0\,\Lambda_{\rm \overline{MS}}$.
This quantity enters $\Delta Z^{\mathcal{S}}(M)$ in  (\ref{RGI2}) via the corresponding coupling
$g^{\mathcal{S}}(M)$ (for details see~\cite{Gockeler:2010yr}). 
We choose $r_0\,\Lambda_{\rm \overline{MS}}=0.700$~\cite{QCDSF:2013} 
and $0.789$~\cite{Fritzsch:2012wq} in order to test the influence
of $r_0\,\Lambda_{\rm \overline{MS}}$.

The fit procedure as sketched above has quite a few degrees of freedom
and it is essential to investigate their influence carefully.
A criterion for the choice of the minimal value of $p^2$ is provided
by the breakdown of perturbation theory at small momenta. The data
suggest~\cite{Gockeler:2010yr} that we are on the 'safe side' when choosing
$p^2_{\min} = 10 \, \mbox{GeV}^2$. 
As the upper end of the fit
interval we take the maximal available momentum at given coupling
$\beta$. 

\vspace{2mm}
\noindent Other important factors are
\vspace{-3mm}
\begin{itemize}

\item {\bf Type of subtraction: } As discussed above, the procedure of the
one-loop subtraction is not unique. We consider the two choices
({\bf s}) and ({\bf m}) with either bare $g$ or
boosted coupling $g_B$. 

\item {\bf Selection of hypercubic invariants:}
For the quality of the fit 
it is essential to have an appropriate description of
the lattice artifacts which remain after subtraction. 
This is connected to the question whether the $a^2$ subtraction has 
been sufficient to subtract (almost) all $a^2$ artifacts. Therefore, we perform
fits with various combinations of structures in (\ref{struc1}).
One should mention that the concrete optimal (i.e.\ minimal) set of  $c_i$ depends
strongly on the momenta of the available Monte Carlo data - momenta close to the diagonal 
in the Brillouin zone require fewer structures to be fitted than far off-diagonal ones.

\end{itemize}

As discussed in detail in~\cite{Constantinou:2013ada} we are not able to find
one combination of the investigated subtraction types and parameter sets $\{c_i\}$ which is
superior to all others. This is partly due to the fact that the available data
set is almost diagonal in the momentum components. Therefore, we rely on our experience, 
which suggests, e.g., the use of the boosted coupling $g_B$ in the 
subtraction prodecure. One further important consideration is the comparison with the
results based on the complete one-loop subtraction 
which serve as benchmarks. In the end, we favor the simple subtraction ({\bf s}) with
boosted coupling $g_B$ using all parameters $c_i$ in the ansatz (\ref{struc1}).

\begin{figure*}[!htb]
  \begin{center}
     \begin{tabular}{lcr}
        \includegraphics[scale=0.5,clip=true]
         {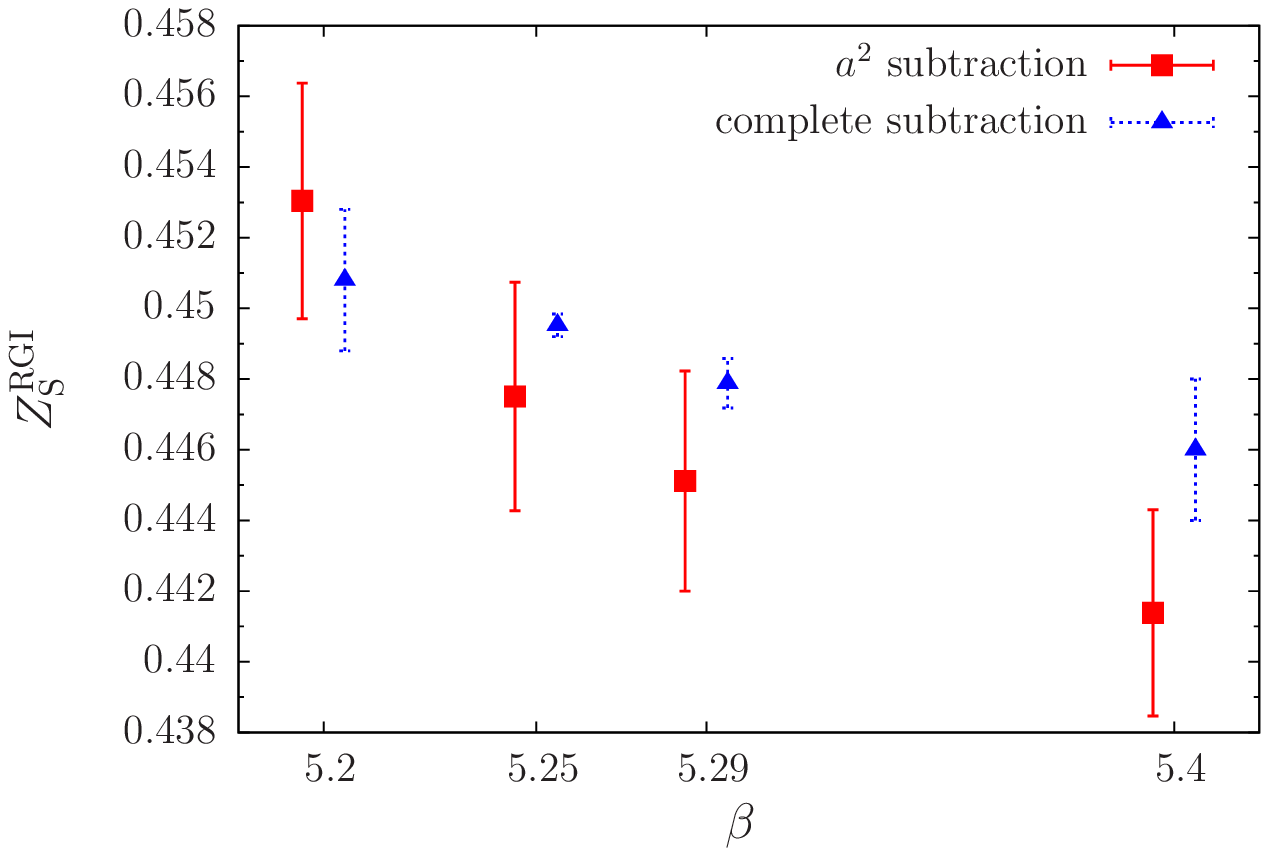}
&~\hspace{1cm}~&
       \includegraphics[scale=0.5,clip=true]
         {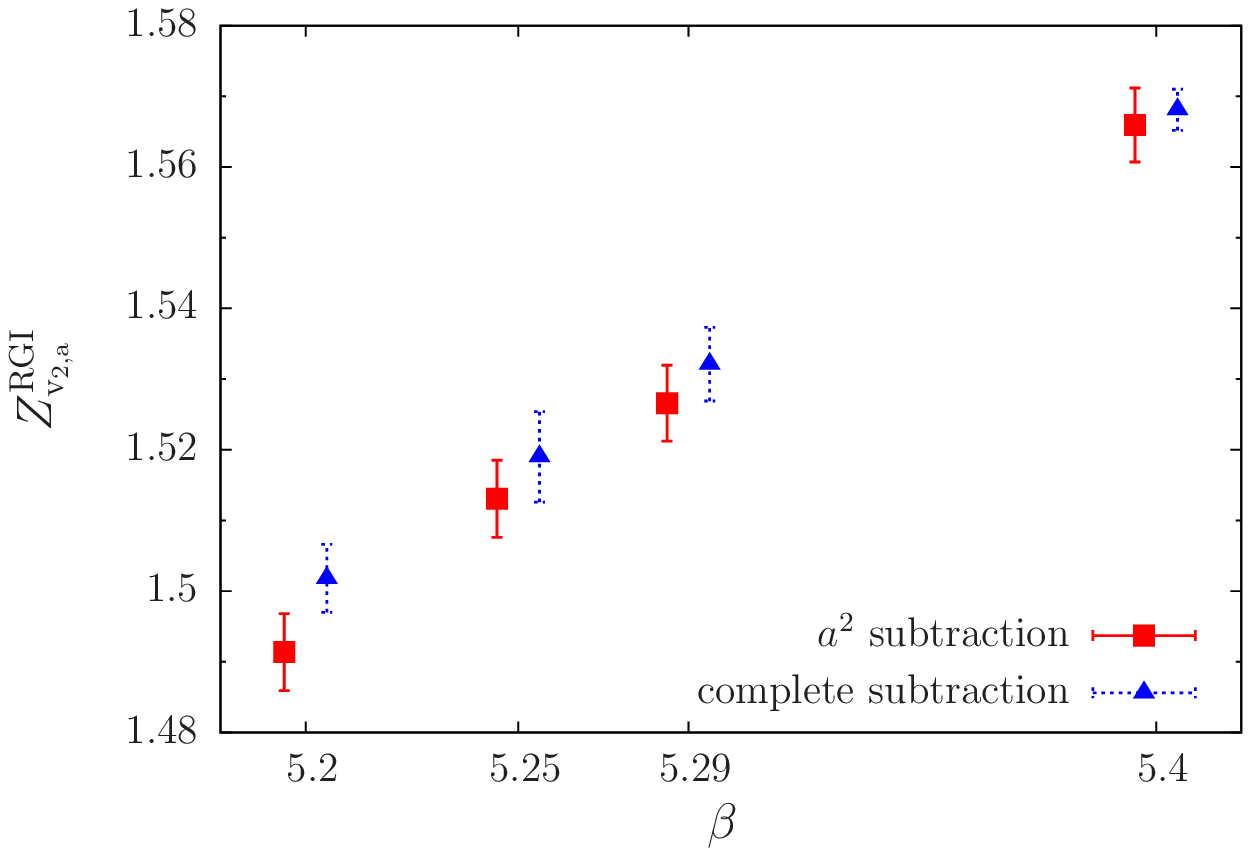}
     \end{tabular}
  \end{center}\vspace{-0.3cm}
  \caption[]{\label{fig:pSva}$Z_S^{\rm RGI}$  (left) and $Z_{v_{2,a}}^{\rm RGI}$ (right) at $r_0\,\Lambda_{\rm \overline{MS}}=0.700$ 
as a function of $\beta$ using
all $c_i$ compared to the complete one-loop 
subtraction.}
\end{figure*}

In Fig.~\ref{fig:pSva} we show as examples the corresponding results for the  
operators $\mathcal{O}^{S}$  and $\mathcal{O}^{v_{2,a}}$ 
compared to the results obtained by the complete one-loop subtraction. 

The final renormalization factors are collected in Table~\ref{tabZRGIm}
\renewcommand*\arraystretch{1.5}
\begin{table*}[!htb]
\begin{center}
  \begin{tabular}{||c|c|c|c|c|c||}
\hline
Op. & $r_0\,\Lambda_{\rm \overline{MS}}$ & $Z^{\rm RGI}\vert_{\beta=5.20}$ & $Z^{\rm RGI}\vert_{\beta=5.25}$ &  $Z^{\rm RGI}\vert_{\beta=5.29}$ & $Z^{\rm RGI}\vert_{\beta=5.40}$ \\ 
\hline
$\mathcal{O}^S $&$ 0.700 $&$  0.4530(34) $&$  0.4475(33) $&$  0.4451(32) $&$  0.4414(30)$\\
$  $&$ 0.789 $&$  0.4717(44) $&$  0.4661(65) $&$  0.4632(54) $&$  0.4585(27)$\\
\hline
$\mathcal{O}^V $&$ 0.700 $&$  0.7163(26) $&$  0.7253(26) $&$  0.7308(25) $&$  0.7451(24)$\\
$  $&$ 0.789 $&$  0.7238(72) $&$  0.7319(94) $&$  0.7365(99) $&$  0.7519(50)$\\
\hline
$\mathcal{O}^A $&$ 0.700 $&$  0.7460(41) $&$  0.7543(40) $&$  0.7590(39) $&$  0.7731(37)$\\
$  $&$ 0.789 $&$  0.7585(46) $&$  0.7634(77) $&$  0.7666(81) $&$  0.7805(30)$\\
\hline
$\mathcal{O}^T $&$ 0.700 $&$  0.8906(43) $&$  0.9036(42) $&$  0.9108(41) $&$  0.9319(39)$\\
$  $&$ 0.789 $&$  0.8946(85) $&$  0.9041(111) $&$  0.9075(120) $&$  0.9316(49)$\\
\hline
$\mathcal{O}^{v_{2,a}} $&$ 0.700 $&$  1.4914(55) $&$  1.5131(55) $&$  1.5266(54) $&$  1.5660(53)$\\
$  $&$ 0.789 $&$  1.4635(108) $&$  1.4776(112) $&$  1.4926(90) $&$  1.5397(58)$\\
\hline
$\mathcal{O}^{v_{2,b}} $&$ 0.700 $&$  1.5061(37) $&$  1.5218(37) $&$  1.5329(36) $&$  1.5534(35)$\\
$  $&$ 0.789 $&$  1.4601(151) $&$  1.4727(206) $&$  1.4863(165) $&$  1.5115(140)$\\
\hline
\end{tabular}
 \end{center}
  \caption{\label{tabZRGIm}$Z^{\rm RGI}$ values using the subtraction ({\bf s}) with $g_B$. The 
errors are obtained from the nonlinear fit procedure.}
\end{table*}
using the two different $r_0\,\Lambda_{\rm \overline{MS}}$ values $0.700$
and $0.789$.
This shows  the influence of the choice of 
$r_0\,\Lambda_{\rm \overline{MS}}$ (depending on the anomalous dimension
of the operator). For the investigated operators and
$\beta$ values we find for the relative differences
of the $Z^{\rm RGI}$ 
\begin{equation}
 \delta Z^{\rm RGI} = \Bigg|\frac{Z^{\rm RGI}_{r_0\,\Lambda_{\rm \overline{MS}}=0.700}-
Z^{\rm RGI}_{r_0\,\Lambda_{\rm \overline{MS}}=0.789}} 
{Z^{\rm RGI}_{r_0\,\Lambda_{\rm \overline{MS}}=0.700}}\Bigg| \lesssim 0.04\,.
\end{equation}
The $Z$ factors of the local (one-link) operators differ with $1\,\%$ ($2\,\%$)
from the corresponding results obtained via the complete one-loop subtraction.

From the present investigation we conclude:
The alternatively
proposed 'reduced' subtraction algorithm can be used for the
determination of the renormalization factors if the complete
subtraction method is not available.
Possible applications could be $Z$ factors for $N_f = 2+1$ 
calculations with more complicated fermionic and gauge actions
where one-loop results to order $a^2$ are available
(for the fermionic SLiNC action with improved 
Symanzik gauge action see Ref.~\cite{Skouroupathis:2010zz}).

\begin{acknowledgments}
This work has been supported in part by the DFG under contract 
SFB/TR55 (Hadron Physics from Lattice QCD) and by the EU grant 283286 (HadronPhysics3).
M. Constantinou, M. Costa and H. Panagopoulos acknowledge support from
the Cyprus Research Promotion Foundation under Contract No.
TECHNOLOGY/$\Theta$E$\Pi$I$\Sigma$/ 0311(BE)/16.
\end{acknowledgments}


\begin{thebibliography}{99}
\bibitem{Martinelli:1994ty}
  G.~Martinelli, C.~Pittori, C.~T.~Sachrajda, M.~Testa and A.~Vladikas,
 Nucl.\ Phys.\ B {\bf 445} (1995) 81
[\href{http://arxiv.org/abs/hep-lat/9411010}{arXiv:hep-lat/9411010}].


\bibitem{Constantinou:2013ada}
  M.~Constantinou, M.~Costa, M.~G\"ockeler, R.~Horsley, H.~Panagopoulos, H.~Perlt, P.~E.~L.~Rakow, G.~Schierholz and A.~Schiller,
  Phys.\ Rev.\  D {\bf 87} (2013) 096019 
[\href{http://arxiv.org/abs/1303.6776}{arXiv:1303.6776[hep-lat]}].


\bibitem{Gockeler:2010yr}
  M.~G\"ockeler, R.~Horsley, Y.~Nakamura, H.~Perlt, D.~Pleiter, P.~E.~L.~Rakow, A.~Sch\"afer,
G.~Schierholz, A.~Schiller, H.~St\"uben and J.~M.~Zanotti,
(QCDSF/UKQCD Collaborations),
  Phys.\ Rev.\ D {\bf 82} (2010) 114511
   [Erratum-ibid.\ D {\bf 86} (2012) 099903]
[\href{http://arxiv.org/abs/1003.5756}{arXiv:1003.5756[hep-lat]}].

\bibitem{Constantinou:2009tr}
  M.~Constantinou, V.~Lubicz, H.~Panagopoulos and F.~Stylianou,
 JHEP {\bf 0910} (2009) 064
[\href{http://arxiv.org/abs/0907.0381}{arXiv:0907.0381[hep-lat]}].

\bibitem{Gockeler:1996mu}
  M.~G\"ockeler, R.~Horsley, E.-M.~Ilgenfritz, H.~Perlt, P.~E.~L.~Rakow, G.~Schierholz and A.~Schiller,
 Phys.\ Rev.\ D {\bf 54} (1996) 5705
[\href{http://arxiv.org/abs/hep-lat/9602029}{arXiv:hep-lat/9602029}].

\bibitem{Bali:2012qs}
  G.~S.~Bali, P.~C.~Bruns, S.~Collins, M.~Deka, B.~Gl\"a\ss le, M.~G\"ockeler,
L.~Greil, T.R.~Hemmert, R.~Horsley, J.~Najjar, Y.~Nakamura, A.~Nobile,
D.~Pleiter, P.~E.~L.~Rakow, A.~Sch\"afer, R.~Schiel, G.~Schierholz, A.~Sternbeck
and J.~M.~Zanotti (QCDSF Collaboration),
 Nucl.\ Phys.\ B {\bf 866} (2013) 1
[\href{http://arxiv.org/abs/1206.7034}{arXiv:1206.7034[hep-lat]}].

 
\bibitem{Alexandrou:2012mt}
  C.~Alexandrou, M.~Constantinou, T.~Korzec, H.~Panagopoulos and F.~Stylianou,
Phys.\ Rev.\ D {\bf 86} (2012) 014505
[\href{http://arxiv.org/abs/1201.5025}{arXiv:1201.5025[hep-lat]}].


\bibitem{QCDSF:2013}
QCDSF collaboration, in preparation.

\bibitem{Fritzsch:2012wq}
  P.~Fritzsch, F.~Knechtli, B.~Leder, M.~Marinkovic, S.~Schaefer, R.~Sommer and F.~Virotta (ALPHA Collaboration),
  Nucl.\ Phys.\ B {\bf 865} (2012) 397
[\href{http://arxiv.org/abs/1205.5380}{arXiv:1205.5380[hep-lat]}].




\bibitem{Skouroupathis:2010zz} 
  A.~Skouroupathis and H.~Panagopoulos,
 PoS LATTICE {\bf 2010}, 240 (2010).



\end{thebibliography}
\end{document}